\documentclass[epjST]{svjour}
\usepackage{graphics} 
\usepackage{epsfig} 
\usepackage{mathptmx} 
\usepackage{amsmath} 
\usepackage{amssymb}  
\usepackage{graphicx}
\usepackage{dcolumn}
\usepackage{bm}
\usepackage{comment}
\pdfoutput=1

\begin{document}
\title{Symbolic local information transfer}
\subtitle{}
\author{Kohei Nakajima\inst{1,2}\fnmsep\thanks{\email{jc\_mc\_datsu@yahoo.co.jp}} \and Taichi Haruna\inst{3}}
\institute{Department of Informatics, University of Zurich, Andreasstrasse 15, 8050 Zurich, Switzerland \and Department of Mechanical and Process Engineering, ETH Zurich, Leonhardstrasse 27, 8092 Zurich, Switzerland \and Department of Earth \& Planetary Sciences, Kobe University, 1-1 Rokkodaicho, Nada, Kobe 657-8501, Japan}
\abstract{
Recently, the permutation-information theoretic approach has been used in a broad range of research fields.  
In particular, in the study of high-dimensional dynamical systems, it has been shown that this approach can be effective in characterizing global properties, including the complexity of their spatiotemporal dynamics. 
Here, we show that this approach can also be applied to reveal local spatiotemporal profiles of distributed computations existing at each spatiotemporal point in the system. J. T. Lizier {\it et al.} have recently introduced the concept of local information dynamics, which consists of information storage, transfer, and modification. This concept has been intensively studied with regard to cellular automata, and has provided quantitative evidence of several characteristic behaviors observed in the system. In this paper, by focusing on the local information transfer, we demonstrate that the application of the permutation-information theoretic approach, which introduces natural symbolization methods, makes the concept easily extendible to systems that have continuous states. We propose measures called {\it symbolic local transfer entropies}, and apply these measures to two test models, the coupled map lattice (CML) system and the Bak-Sneppen model (BS-model), to show their relevance to spatiotemporal systems that have continuous states. In the CML, we demonstrate that it can be successfully used as a spatiotemporal filter to stress a coherent structure buried in the system. In particular, we show that the approach can clearly stress out defect turbulences or Brownian motion of defects from the background, which gives quantitative evidence suggesting that these moving patterns are the information transfer substrate in the spatiotemporal system. We then show that these measures reveal qualitatively different properties from the conventional approach using the sliding window method, and are also robust against external noise. In the BS-model, we demonstrate that these measures can provide novel insight to the model, featuring how symbolic local information transfer is related to 
the dynamical properties of the elements involved in a spatiotemporal dynamics.
} 
\maketitle
\section{Introduction}
\label{intro}
Understanding the properties of a spatiotemporal system that consists of a large number of interacting elements has been an important research topic for several decades. In such a system, it has often been suggested that local interactions in space can give rise to global spatiotemporal patterns, and huge amounts of effort have been devoted to characterizing the behaviors of the spatiotemporal dynamics of the system \cite{Wolfram,Langton,Bak,Kauffman,Kaneko0,Kuramoto}. Information theoretic approach is one of the main methods for analyzing such systems. For example, mutual information (MI) has been intensively used to characterize the class of spatiotemporal dynamics revealed in cellular automata \cite{Langton}, the order-chaos transition in random boolean networks \cite{boolean}, or the interaction regimes in coupled map lattice (CML) systems \cite{Kaneko3,Schreiber2}. Also, the diversity of information processing capacity of spatiotemporal dynamics has been characterized in terms of information structures using excess entropy (or, the complexity-entropy diagram) \cite{excess}. Although these analyses reveal important statistical properties of the global behavior of spatiotemporal dynamics, they often miss capturing their spatiotemporal profile or the structure of the dynamics, namely the characteristics distributed at each spatiotemporal point in the system. 

Recently, J. T. Lizier {\it et al.} introduced the concept of information dynamics for the analysis of distributed computations in spatiotemporal dynamics, which addresses this inadequacy \cite{Lizier5}. The information dynamics consist of the fundamental operations of information storage, transfer, and modification, where these operations are quantified on a local scale in space and time. 
They have formalized these operations in terms of information theoretic quantities for spatiotemporal dynamics, and they are called {\it local information storage} \cite{Lizier3}, {\it local information transfer} \cite{Lizier1}, and {\it local information modification} \cite{Lizier4}, respectively. These measures have been intensively studied with regard to elementary cellular automata (ECA), and have provided quantitative evidence of several characteristic behaviors observed in the system. For example, they introduced the measure of local information transfer based on a measure called {\it transfer entropy} (TE) \cite{Schreiber}, and showed that a typical emergent agent in ECA, known as {\it glider} and long considered an information transfer substrate from the observation of the spatiotemporal pattern of ECA, was characterized as such in a quantitative manner with an information theoretic term \cite{Lizier1}. 

Our aim in this paper is to provide an extension to make this framework feasible for the spatiotemporal system that has continuous states. This is achieved by introducing a natural symbolization technique for a continuous time series, which uses local orderings (permutations) of the values as associated states \cite{Bandt1}. This approach is called the permutation-information theoretic approach, and it was first introduced as a permutation version of Shannon entropy, known as {\it permutation entropy} (PE) \cite{Bandt1}. It has been shown to be equivalent to the original if both are considered as the rates for a stationary time series under a certain assumption \cite{Amigo1,Amigo2,Kohei1,Amigo3,Keller1,Keller2}, and has been extensively used because of its ease of implementation, its computational efficiency, and its robustness against external noise \cite{Amigo1,PE1,PE2,PE3}.
Several measures have recently been proposed based on this technique \cite{Kohei1,Kohei2,Kohei4,Amigo4,PEE1,PEE2,Amigo5}. 
In this paper, we only focus on the concept of local information transfer for spatiotemporal dynamics and propose measures in the realm of {\it symbolic local information transfer} based on the permutation versions of TE. In order to demonstrate their relevance to the spatiotemporal system that has continuous states, we apply these measures to the CML system and the Bak-Sneppen model (BS-model), and show that they can provide the characteristic spatiotemporal profile in each system.

This paper is organized as follows.
In section \ref{sec1}, we give a brief overview of information theoretic preliminaries and introduce the measure of TE. We then explain the permutation expression of the measure adopted in this paper. In section \ref{sec3}, we explain how to make the measure {\it local} to be applied for the spatiotemporal system by reviewing the procedure introduced in \cite{Lizier1} and propose the measures of {\it symbolic local transfer entropies}. In section \ref{sec4}, we apply these measures to the CML and the BS-model and analyze their spatiotemporal profile in detail. In particular, for the CML, we compare the performance of the measures with conventional techniques based on sliding windows, and also investigate the robustness of the measures against external noise. For the BS-model, we provide some explanations of the local information transfer profile, which are specifically useful for understanding the properties of this model. Finally, in section \ref{sec6}, we give conclusions and discuss future extension scenarios for these measures.  

\section{Information transfer and its permutation expression}
\label{sec1}
We begin with a brief overview of the background of the information theoretic concepts required in this paper \cite{Cover}.
The uncertainty associated with the state $x$ of a random variable $X$ following a probability distribution $p(x)$ is given by the Shannon entropy:
\begin{equation}
H(X) = - \sum_{x} p(x) \log_{2} p(x),
\end{equation}
where the base of the logarithm is taken as $2$ throughout this paper; therefore, the unit of all measures we present in this paper is unified as bit. 
The MI between two processes $X$ and $Y$ measures the average information gained about $X$ by knowing the process of $Y$, or vice versa, as follows:
\begin{equation}
M_{XY} = \sum_{x, y} p(x, y) \log \frac{p(x, y)}{p(x) p(y)},
\end{equation}
where $p(x, y)$ is a joint probability distribution of processes $X$ and $Y$ \cite{Cover}.
For statistically independent distributions, 
$p(x, y) = p(x) p(y)$ and $M_{XY} =0$.
If statistical dependencies exist, $M_{XY} > 0$.
MI is a fundamental measure in information theory and is used to evaluate an association between two or more processes, which naturally encompass linear and nonlinear dependencies.
However, as can be seen from the equation, MI is intrinsically symmetric under the exchange of the two processes $X$ and $Y$, which means that MI does not contain any directional information.

In this paper, we focus on the notion of information transfer, which requires capturing both directional and dynamic relations from an information source to a receiver.
Therefore, MI is insufficient for this purpose. 
There are several measures proposed to address these inadequacies (see \cite{Schreiber,momentary,DI1,DI2,DI3}, for example). 
Our focus here is on the measure known as {\it transfer entropy} (TE) proposed by Schreiber \cite{Schreiber}. 
TE is a measure of the information transfer from the driving system ($Y$)
to the responding system ($X$). 
Let us write $x_{t}$ and $y_{t}$ for the values of two temporal processes $X_{t}$ and $Y_{t}$, respectively.
TE essentially quantifies the deviation from the generalized Markov property: 
$p(x_{t+1} | x_{t}^{(K)}) = p(x_{t+1} | x_{t}^{(K)}, y_{t}^{(L)})$, where $p(x_{t+1} | x_{t}^{(K)})$ denotes a transition probability from $x_{t}^{(K)}$ to $x_{t+1}$, and $K, L$ are the length of the delay embedding vectors.
If the deviation from a generalized Markov process is small,
then the state $y_{t}^{(L)}$ can be assumed to have little relevance
to the transition probabilities from $x_{t}^{(K)}$ to $x_{t+1}$. If the deviation
is large, however, then the assumption of a Markov process
is not valid. The incorrectness of the assumption can be
expressed by the TE, formulated as a specific
version of the Kullback-Leibler entropy \cite{Cover,Schreiber}:
\begin{equation}\label{TE}
TE_{Y \rightarrow X} = \sum_{x_{t+1}, x_{t}^{(K)}, y_{t}^{(L)}} p(x_{t+1}, x_{t}^{(K)}, y_{t}^{(L)}) \log \frac{p(x_{t+1} | x_{t}^{(K)}, y_{t}^{(L)})}{p(x_{t+1} | x_{t}^{(K)})},
\end{equation}
where the index $TE_{Y \rightarrow X}$ indicates the influence of $Y$ on $X$, and can thus be used to detect the directed information
transfer. In other words, TE measures how well
we can predict the transition of system $X$ by knowing system $Y$. TE is non-negative, and any information transfer between the
two variables results in $TE_{Y \rightarrow X}>0$. If the state $y_{t}^{(L)}$ has no
influence on the transition probabilities from $x_{t}^{(K)}$ to $x_{t+1}$, or if
the two time series are completely synchronized, then $TE_{Y \rightarrow X}=0$.
TE can also be denoted as follows:
\begin{equation}
TE_{Y \rightarrow X} = -H(X_{t+1}, X_{t}, Y_{t}) + H(X_{t}, Y_{t}) + H(X_{t+1}, X_{t}) -H(X_{t}),
\end{equation}
where $H(X_{t+1}, X_{t}, Y_{t})$, $H(X_{t}, Y_{t})$, and $H(X_{t+1}, X_{t})$ are joint entropies of corresponding temporal processes.

Our aim in this paper is to apply information theoretic measures to a spatiotemporal time series. 
As explained above, the entropies are functionals of probability distributions, and there are variations of methods proposed to estimate the distributions from an obtained time series \cite{TE18}. One popular class of estimators divides the given range of state space into a set of partitions, which is often referred to as binning.
On the other hand, PE quantifies the uncertainty of the local orderings
of values, unlike the usual entropy, which quantifies that of the values themselves. 
This approach introduces a natural symbolization technique for continuous states as we will see later.
In spite of the differences between the procedures, it was proven that the
PE rate is equal to the usual entropy rate for any finite-alphabet stationary stochastic process \cite{Amigo1,Kohei1}. It was also shown that PE is robust to noise, which is common in a real-world time series \cite{Bandt1}. PE is derived from this permutation partition.
Let $x_{t}^{(L)}$ represent
an $L$ dimensional embedding vector from the obtained time series $x'_{t}$, 
and $\hat{x}_{t}^{(L)}$ 
be a sequence of numbers representing the orderings of $x_{t}^{(L)}$. 
Based on the permutations of the values, $\hat{x}_{t}^{(L)}$ is
generated as follows: 
$x_{t}^{(L)} = ( x'_{t}, x'_{t-1}, ..., x'_{t-(L-1)} )$, and 
are arranged in ascending order, 
$ x'_{t-(o_{t} (1) -1)} \leq  x'_{t-(o_{t} (2) -1)} \leq ... \leq x'_{t-( o_{t} (L) -1)} $.
A symbol is thus defined as 
$\hat{x}_{t}^{(L)} \equiv (o_{t} (1), o_{t} (2), ..., o_{t} (L) ) \in \hat{X}_{t}$, 
where $\hat{X}_{t}$ is the set of symbols generated in the temporal process $X_{t}$.
Based on the generated symbols $\hat{x}_{t}^{(L)}$, PE is expressed
as:
\begin{equation}
H(\hat{X}_{t}) = - \sum_{\hat{x}_{t}^{(L)}} p(\hat{x}_{t}^{(L)}) \log p(\hat{x}_{t}^{(L)}),
\end{equation}
where $p(\hat{x}_{t}^{(L)})$ is the probability of the occurrence of $\hat{x}_{t}^{(L)}$ in the set of symbols $\hat{X}_{t}$. 

Similarly to TE, its permutation version is proposed in \cite{Symbolic} and is called {\it symbolic
transfer entropy} (STE), expressed as:
\begin{equation}
ST_{Y \rightarrow X} = -H(\hat{X}_{t+1}, \hat{X}_{t}, \hat{Y}_{t}) + H(\hat{X}_{t}, \hat{Y}_{t}) + H(\hat{X}_{t+1}, \hat{X}_{t}) -H(\hat{X}_{t}),
\end{equation}
where $H(\hat{X}_{t+1}, \hat{X}_{t}, \hat{Y}_{t})$, $H(\hat{X}_{t}, \hat{Y}_{t})$, and $H(\hat{X}_{t+1}, \hat{X}_{t})$ are joint entropies of the set of symbols in the corresponding temporal processes, and in terms of the probability distribution functions as:
\begin{equation}\label{STE}
ST_{Y \rightarrow X} = \sum_{ \hat{x}_{t+1}^{(M)}, \hat{x}_{t}^{(K)}, \hat{y}_{t}^{(L)} } p(\hat{x}_{t+1}^{(M)}, \hat{x}_{t}^{(K)}, \hat{y}_{t}^{(L)} ) \log \frac{p(\hat{x}_{t+1}^{(M)} | \hat{x}_{t}^{(K)}, \hat{y}_{t}^{(L)} )}{p(\hat{x}_{t+1}^{(M)} | \hat{x}_{t}^{(K)} )},
\end{equation}
where $M, K,$ and $L$ are the length of the embedding vectors, and the index $ST_{Y \rightarrow X}$ indicates the influence of $Y$ on $X$. 

Note that Eq. (\ref{STE}) is not completely analogous to the original expression of TE (Eq. (\ref{TE})) because of the embedding vector $M$ defined for $\hat{x}_{t+1}^{(M)}$ in the set of symbols $\hat{X}_{t+1}$. 
This setting introduces the overlap of the symbols between $\hat{x}_{t+1}^{(M)}$ ($\hat{x}_{t+1}^{(M)} \in \hat{X}_{t+1}$) and $\hat{x}_{t}^{K}$ ($\hat{x}_{t}^{K} \in \hat{X}_{t}$), which also introduces the intrinsic bias to the measure \cite{Kugiumtzis}. 
In order to overcome this effect of bias, the improved version of STE was proposed in \cite{Kugiumtzis}, called {\it transfer entropy on rank vectors} (TERV). 
The basic concept of TERV is to use further timesteps ahead to avoid this overlap of symbols, namely a symbol $\hat{x}_{t+1}^{(M)}$ is generated from the obtained time series vector, $(x'_{t+1}, x'_{t+2}, ..., x'_{t+1+(M-1)})$, instead of the original one, $(x'_{t+1}, x'_{t}, ..., x'_{t+1-(M-1)})$, while the other symbols $\hat{x}_{t}^{(K)}$ and $\hat{y}_{t}^{(L)}$ are generated by the original procedure. We distinguish the former denoting the symbol as $\hat{x}_{t+1}^{(M), ahead}$ from the latter.
Accordingly, TERV can be expressed as:
\begin{equation}\label{TERV}
TERV^{M}_{Y \rightarrow X} = \sum_{ \hat{x}_{t+1}^{(M), ahead}, \hat{x}_{t}^{(K)}, \hat{y}_{t}^{(L)} } p(\hat{x}_{t+1}^{(M), ahead}, \hat{x}_{t}^{(K)}, \hat{y}_{t}^{(L)} ) \log \frac{p(\hat{x}_{t+1}^{(M), ahead} | \hat{x}_{t}^{(K)}, \hat{y}_{t}^{(L)} )}{p(\hat{x}_{t+1}^{(M), ahead} | \hat{x}_{t}^{(K)} )}.
\end{equation}
In this paper, we use the special case of TERV in which the embedding vector $M$ is set to 1 expressed as:
\begin{equation}
TERV^{1}_{Y \rightarrow X} = \sum_{ \hat{x}_{t+1}, \hat{x}_{t}^{(K)}, \hat{y}_{t}^{(L)} } p(\hat{x}_{t+1}, \hat{x}_{t}^{(K)}, \hat{y}_{t}^{(L)} ) \log \frac{p(\hat{x}_{t+1} | \hat{x}_{t}^{(K)}, \hat{y}_{t}^{(L)} )}{p(\hat{x}_{t+1} | \hat{x}_{t}^{(K)} )}.
\end{equation}
This setting allows us to combine the tuple $(\hat{x}_{t+1}^{(M=1), ahead}, \hat{x}_{t}^{(K)}, \hat{y}_{t}^{(L)})$ into $(\hat{x}_{t+1}^{(K+1)}, \hat{y}_{t}^{(L)})$ when generating the symbol and calculating the TERV from the obtained time series \cite{Kugiumtzis}, since $\hat{x}_{t+1}$ alone cannot form a permutation ordering. Accordingly, TERV with $M=1$ can be expressed as follows:
\begin{equation}\label{TERV1}
TERV^{1}_{Y \rightarrow X} = -H(\hat{X}_{t+1}, \hat{Y}_{t}) + H(\hat{X}_{t}, \hat{Y}_{t}) + H(\hat{X}_{t+1}) -H(\hat{X}_{t}).
\end{equation}
This forms the permutation expression of the information transfer in this paper \cite{Kohei3,Kohei4}. 
When we actually calculate the value of $TERV^{1}_{Y \rightarrow X}$, we first calculate the joint and single entropies in the right side of Eq. (\ref{TERV1}) from the corresponding probability distributions, namely $p(\hat{x}_{t+1}^{(K+1)}, \hat{y}_{t}^{(L)}), p(\hat{x}_{t}^{(K)}, \hat{y}_{t}^{(L)}), p(\hat{x}_{t+1}^{(K+1)}),$ and $p(\hat{x}_{t}^{(K)})$, with the obtained time series. 

\section{Local information transfer and its permutation expressions for spatiotemporal systems}
\label{sec3}

\begin{figure}[htbp]
	\centerline{\includegraphics[width=5.0in, bb=80 453 498 745]{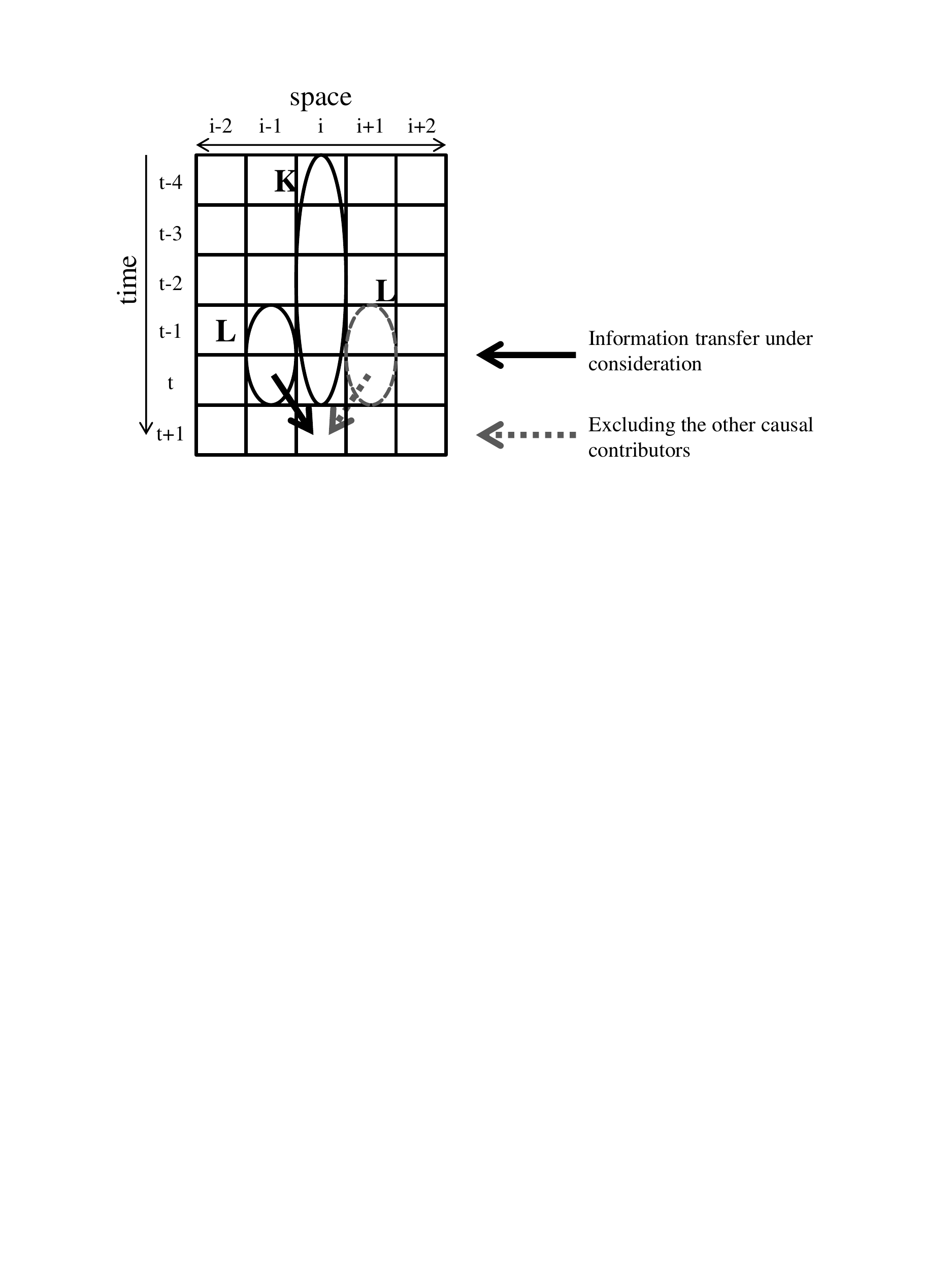}}
	\caption{Schematic showing the local information transfers in a spatiotemporal system. The solid arrow shows the local information transfer from a cell $i-1$ to cell $i$ (LATE) at timestep $t+1$, where the embedding vector of each state, $x_{i-1, t}^{L}$ or $x_{i, t}^{K}$, is $(K, L) = (5, 2)$ in the figure. LCTE is obtained by conditioning out all the other causal contributors (the dashed arrow) to the destination cell $i$. In the CML and the BS-model, since each cell only interacts with its nearest neighboring cells ($i-1, i+1$), to obtain the LCTE from a cell $i-1 (i+1)$, we condition out the contribution from $i+1 (i-1)$. See text for details.}
	\label{schematics}
\end{figure}

In this section, we review the procedure for making TE local. In \cite{Lizier1}, based on TE, Lizier {\it et al.} focused on the fact that, in obtaining the value of TE from the time series, the joint probability $p(x_{t+1}, x_{t}^{(K)}, y_{t}^{(L)})$ is operationally equivalent to the ratio of the count of observations $c(x_{t+1}, x_{t}^{K}, y_{t}^{(L)})$ of state transition tuples $(x_{t+1}, x_{t}^{(K)}, y_{t}^{(L)})$, to the total number of observations $N$ made. In applications to time series, the number of observations is finite, and $p(x_{t+1}, x_{t}^{(K)}, y_{t}^{(L)})$ can be expressed as $p(x_{t+1}, x_{t}^{(K)}, y_{t}^{(L)}) = c(x_{t+1}, x_{t}^{(K)}, y_{t}^{(L)})/N$. Then, TE can be expressed as follows:
\begin{equation}
TE_{Y \rightarrow X} = \frac{1}{N} \sum_{x_{t+1}, x_{t}^{(K)}, y_{t}^{(L)}} (\sum_{a=1}^{c(x_{t+1}, x_{t}^{(K)}, y_{t}^{(L)})} 1) \log \frac{p(x_{t+1} | x_{t}^{(K)}, y_{t}^{(L)})}{p(x_{t+1} | x_{t}^{(K)})}.
\end{equation}
By considering that a double sum running over each actual observation $a$ for each possible tuple observation $(x_{t+1}, x_{t}^{(K)}, y_{t}^{(L)})$ is nothing but a single sum over all $N$ observations, we obtain the following:
\begin{equation}
TE_{Y \rightarrow X} = \frac{1}{N} \sum_{all \ observations} \log \frac{p(x_{t+1} | x_{t}^{(K)}, y_{t}^{(L)})}{p(x_{t+1} | x_{t}^{(K)})}.
\end{equation}
Thus, we can write TE as the global average over {\it local transfer entropy} (LTE), $te_{Y \rightarrow X} (t+1)$, defined as,
\begin{align}
TE_{Y \rightarrow X} =<te_{Y \rightarrow X} (t+1)>, \\
te_{Y \rightarrow X} (t+1) = \log \frac{p(x_{t+1} | x_{t}^{(K)}, y_{t}^{(L)})}{p(x_{t+1} | x_{t}^{(K)})},
\end{align}
where $<X>$ denotes the expectation value of $X$.
Note that LTE can have a negative value. The negative value of LTE means that the sender is misleading about the prediction of the receiver's next state \cite{Lizier1}. 
For a spatiotemporal system, where the senders and receivers are spatially ordered cells, the LTE to cell $i$ from cell $i-j$ at timestep $t+1$ (Fig.\ref{schematics}) can be expressed as:
\begin{equation}\label{apparent}
te(i, j, t+1) = \log \frac{p(x_{i, t+1} | x_{i, t}^{(K)}, x_{i-j, t}^{(L)})}{p(x_{i, t+1} | x_{i, t}^{(K)})}.
\end{equation}
This measure is defined for every spatiotemporal receiver $(i, t)$, forming a spatiotemporal profile for every information direction $j$. Note that $j$ represents the number of cells from the sender to the receiver. For example, if $j=1$ or $-1$, then $te(i, j, t+1)$ expresses the LTE from the spatially neighboring cell to the cell $i$. As we will see later, we only consider spatiotemporal dynamics where each cell $i$ has a local interaction with its neighboring cells, $j=1, -1$, in this paper. Lizier {\it et al.}, then, introduced a variation of the LTE, which considers conditioning out other possible causal information contributors apart from the information source under consideration \cite{Lizier2} (Fig. \ref{schematics}). This measure is called {\it local complete transfer entropy} (LCTE), and expressed as follows:
\begin{align}
te_{c}(i, j, t+1) = \log \frac{p(x_{i, t+1} | x_{i, t}^{(K)}, x_{i-j, t}^{(L)}, v_{i, j, t})}{p(x_{i, t+1} | x_{i, t}^{(K)}, v_{i, j, t})}, \\
v_{i, j, t} = \{ x_{i+q, t}^{(L)} | \forall q: information \ sources, q \neq -j, 0 \},
\end{align}
where $v_{i, j, t}$ is the joint value of the neighborhood of the destination $x_{i, t+1}$, excluding the source for the TE calculation $x_{i-j, t}^{(L)}$ and the previous value of the destination $x_{i, t}^{(K)}$. To explicitly distinguish the LCTE from the original one, the LTE expressed as Eq. (\ref{apparent}) is called {\it local apparent transfer entropy} (LATE), and is denoted as $te_{a}(i, j, t+1)$.
Finally, the summed up version of both LATE and LCTE ($te_{sa}(i, t+1)$ and $te_{sc}(i, t+1)$) over the neighboring cells ($j=1, -1$) are introduced as follows:
\begin{align}
te_{sa}(i, t+1) = \sum_{j : neighborhood} te_{a}(i, j, t+1), \\
te_{sc}(i, t+1) = \sum_{j : neighborhood} te_{c}(i, j, t+1).
\end{align}

In \cite{Lizier1}, these measures were applied to ECA and several interesting properties of its spatiotemporal profile were reported. Our aim here is to extend this framework to spatiotemporal systems which have continuous states. Usually, for continuous time series, the estimation of the probability distributions requires an intensive amount of preconditioning of the data to obtain the appropriate binning of the state or the fine tuning of the parameters of the probability distribution estimators. However, as explained in the previous section, the permutation partitioning can naturally symbolize the continuous states, thus can be effectively applied to the spatiotemporal systems with continuous states. Based on TERV (with $M=1$) explained in the previous section, we can straight-forwardly introduce {\it symbolic local transfer entropies} ({\it symbolic local apparent transfer entropy}, {\it symbolic local complete transfer entropy}, {\it summed symbolic local apparent transfer entropy} (SLA), {\it summed symbolic local complete transfer entropy} (SLC), which are denoted as $ste_{a}(i, j, t+1)$, $ste_{c}(i, j, t+1)$, $ste_{sa}(i, t+1)$, and $ste_{sc}(i, t+1)$, respectively) as follows:
\begin{equation}\label{stea}
ste_{a}(i, j, t+1) = \log \frac{p(\hat{x}_{i, t+1} | \hat{x}_{i, t}^{(K)}, \hat{x}_{i-j, t}^{(L)})}{p(\hat{x}_{i, t+1} | \hat{x}_{i, t}^{K})},
\end{equation}
\begin{equation}\label{stec}
ste_{c}(i, j, t+1) = \log \frac{p(\hat{x}_{i, t+1} | \hat{x}_{i, t}^{(K)}, \hat{x}_{i-j, t}^{(L)}, \hat{v}_{i, j, t})}{p(\hat{x}_{i, t+1} | \hat{x}_{i, t}^{(K)}, \hat{v}_{i, j, t})},
\end{equation}
\begin{equation}\label{vv}
\hat{v}_{i, j, t} = \{ \hat{x}_{i+q, t}^{(L)} | \forall q: information \ sources, q \neq -j, 0 \},
\end{equation}
\begin{equation}
ste_{sa}(i, t+1) = \sum_{j : neighborhood} ste_{a}(i, j, t+1),
\end{equation}
\begin{equation}
ste_{sc}(i, t+1) = \sum_{j : neighborhood} ste_{c}(i, j, t+1).
\end{equation}
In order to calculate the symbolic local transfer entropies, we combine the tuple $(\hat{x}_{i, t+1}, \hat{x}_{i, t}^{(K)})$ into $(\hat{x}_{i, t+1}^{(K+1)})$ and generate the required probability distributions from the obtained time series as explained in the previous section.
In the following sections, we show how these local measures can reveal the spatiotemporal profiles of the spatiotemporal dynamics which have continuous states.
 
\section{Demonstrations}
\label{sec4}
In this section, we demonstrate the power of the symbolic local transfer entropies in analyzing the spatiotemporal profiles of the spatiotemporal dynamics.
As test models of the spatiotemporal dynamics, we selected the CML system \cite{Kaneko2,Kaneko1,Kaneko0} and the BS-model \cite{Bak1993,Flyvbjerg1993}. 
For the CML, the spatiotemporal dynamics is expressed in one-dimensional space, and all the cells interact with the neighboring cells at each timestep with uniformly fixed coupling strength, which is a common setting in spatially distributed dynamical systems.
In contrast, although the BS-model is also expressed in one dimension, the model's form of interaction is different from that of CML. Namely, as explained later, only a cell selected by a certain condition can interact with neighboring cells at each timestep. Due to this specific form of interaction, we can investigate in detail how the local information transfer and the local interaction mechanism of the cells are related to each other in this model. 

\subsection{Symbolic local information transfer in the CML}
The CML used in this paper is expressed as follows:
\begin{align}
x_{i, t+1} &= (1-c)f(x_{i, t}) + \frac{c}{2} (f(x_{i+1, t}) + f(x_{i-1, t})), \\
f(x) &= 1-ax^{2},
\end{align}
where $c$ is the coupling strength set as $c=0.1$, the nonlinear parameter $a$ is selected from the range in $1.5 \leq a \leq 2.0$, $x \in [-1, 1]$ in our experiment and the periodic boundary condition is adopted.
In the CML, it is well known that according to the selection of $c$ and $a$, the system shows qualitatively different spatiotemporal dynamics \cite{Kaneko1,Kaneko0}. The spatiotemporal dynamics has been explained by the characteristic spatial structures, where cells form clusters with various sizes and, in each cluster, show similar behaviors ranging from chaos to oscillatory behaviors. 

\begin{figure}[htbp]
	\centerline{\includegraphics[width=5.0in, bb=16 123 578 678]{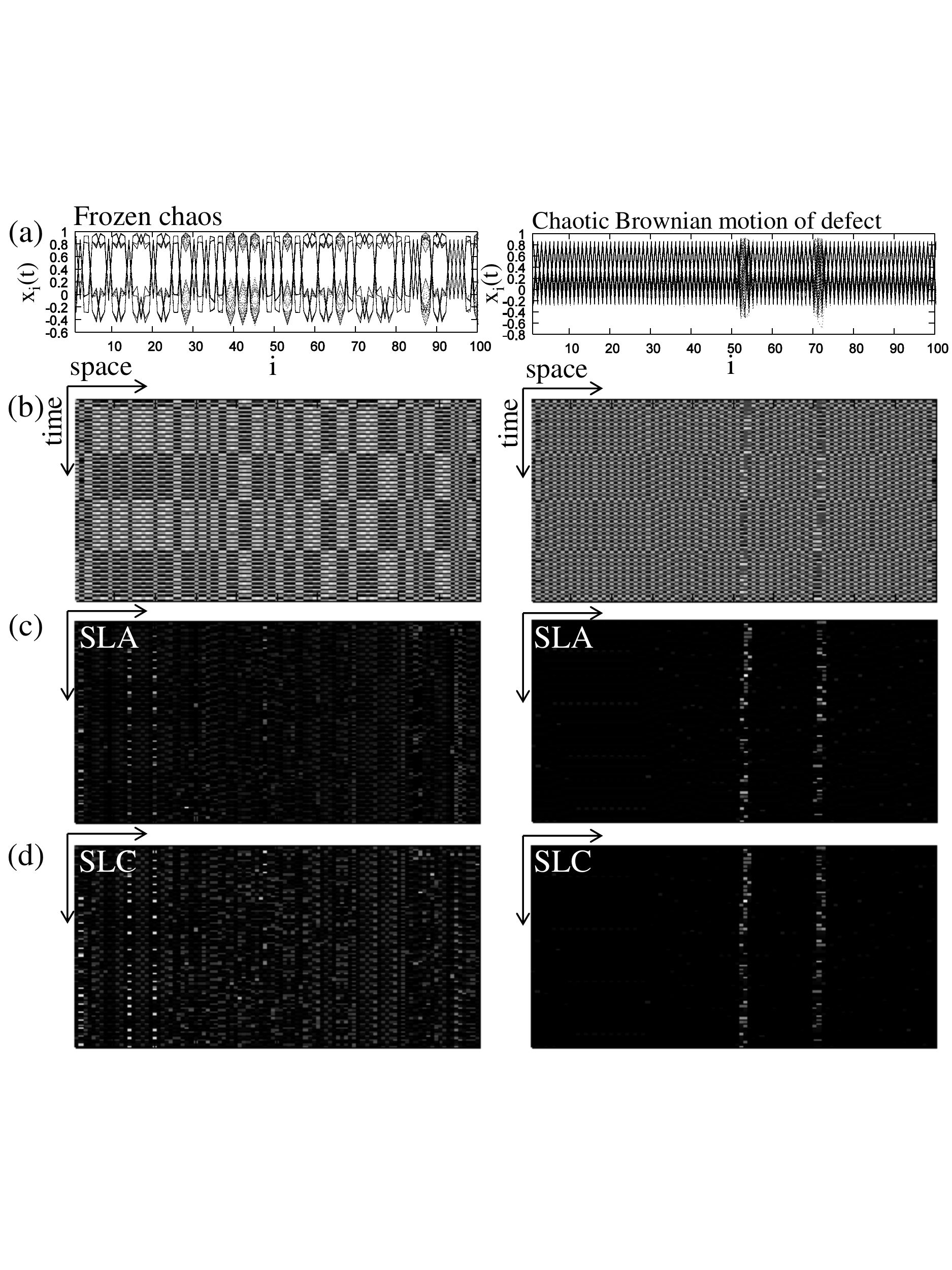}}
	\caption{A typical example of spatiotemporal dynamics and the corresponding local information transfer in the FC (left row) and BD (right row) phases. The system size $N$ is set to 100, and for SLA and SLC, the embedding vectors $(L, K)$ are set to (2, 5). In each phase, we ran the system from a random initial condition for 10000 timesteps, and the initial 1000 timesteps were discarded; the remaining 9000 timesteps were used to calculate the probability distributions for the local information transfers, and the spatiotemporal profile of the last 100 timesteps are shown. (a) $x_{i} (t)$ is overlaid according to the space for each phase. (b) The spatiotemporal dynamics for each phase. (c) The spatiotemporal profile of SLA for each phase. The darker the color, the lower the value of the local information transfer. For the FC (left diagram) and BD (right diagram) phases, the values ranged from 0 to 4 [bit]. (d) The spatiotemporal profile of SLC for each phase. The darker the color, the lower the value of the local information transfer. For the FC (left diagram) and BD (right diagram) phases, the values ranged from 0 to 4 [bit] and from 0 to 5 [bit], respectively.}
	\label{local_phase1}
\end{figure}

\begin{figure}[htbp]
	\centerline{\includegraphics[width=5.0in, bb=16 123 578 678]{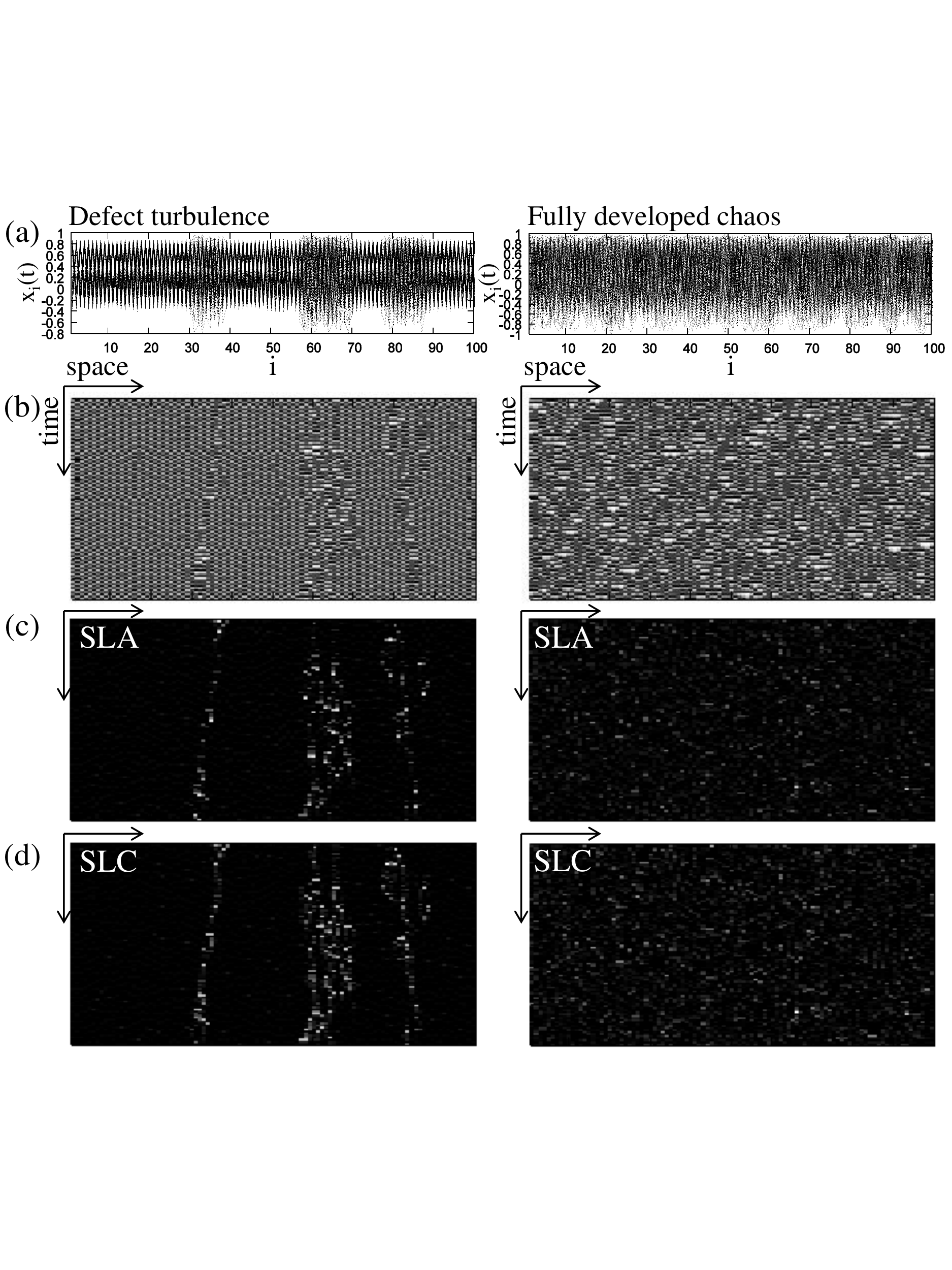}}
	\caption{A typical example of spatiotemporal dynamics and the corresponding local information transfer in the DT (left row) and FDC (right row) phases. The experiment settings are the same as in Fig. \ref{local_phase1}. (a) $x_{i}(t)$ is overlaid according to the space for each phase. (b) The spatiotemporal dynamics for each phase. (c) The spatiotemporal profile of SLA for each phase. The darker the color, the lower the value of the local information transfer. For the DT (left diagram) and FDC (right diagram) phases, the values ranged from 0 to 6 [bit]. (d) The spatiotemporal profile of SLC for each phase. The darker the color, the lower the value of the information transfer. For the DT (left diagram) and FDC (right diagram) phases, the values ranged from 0 to 4.5 [bit] and from 0 to 6 [bit], respectively.}
	\label{local_phase2}
\end{figure}

For example, if the nonlinear parameter $a$ is less than around 1.55, cells form fixed clusters of various sizes according to the initial states selected, and in some clusters show chaotic behaviors (Fig.\ref{local_phase1} (a), (b) left figure). Due to this characteristic, this form of spatiotemporal dynamics is called {\it Frozen chaos} (FC). By increasing the degree of nonlinearity, larger clusters become unstable, and cells start to form smaller clusters. In these clusters, the chaotic behaviors observed in FC are suppressed, and as a result, various oscillatory behaviors are found in each cluster. The characteristic of the size and this oscillatory behavior depend on the setting of $a$ and $c$. Due to this cluster formation process, this form of spatiotemporal dynamics is called {\it Pattern selection} (PS). If we increase the degree of nonlinearity further, the clusters become smaller and stable, but in specific coupling strengths, because several cells miss forming a cluster, the defected formation of clusters start to move around the space randomly in a Brownian manner (Fig. \ref{local_phase1} (a), (b) right figure). This type of spatiotemporal dynamics is called {\it Chaotic Brownian motion of defect} (BD) and can be especially found in relatively smaller value of coupling strengths. Further increase in nonlinearity makes the clusters more unstable, so that both coherent and chaotic spatiotemporal patterns co-exist in the system. In this phase, we can often find {\it Defect turbulence} (DT), where the irregular motion of defects starts to emerge spontaneously (Fig. \ref{local_phase2} (a), (b) left figure). And finally, when the degree of nonlinearity is increased further, the system reaches the state called {\it Fully developed chaos} (FDC), where all the cells reveal chaotic behaviors (Fig. \ref{local_phase2} (a), (b) right figure).

Now, we analyzed these spatiotemporal dynamics in terms of the symbolic local information transfer profile. As explained in the previous section, the symbolic local information transfer values are sometimes negative. In those cases, we set the value to 0 throughout the analyses in this subsection. In addition, we especially focus on the measures SLA and SLC throughout this subsection, but we would like to note that the results using the other symbolic local transfer entropies were qualitatively the same with those reported below. 

Figure \ref{local_phase1} shows the results for SLA and SLC in FC and BD. In FC, since each cell belongs to the fixed clusters, according to the dynamics (chaos or nearly oscillatory behavior) in each cluster, SLA and SLC also reveal a specific profile for each cell (Fig. \ref{local_phase1} (c), (d); left row). The results depend on the dynamics of the cluster that the cell belongs to and the dynamics of the neighboring cluster of the cell, in the degree of how well the sender cell can predict the dynamics of the receiver cell. From this local information transfer profile, we can infer that the spatio-structure of the dynamics is fixed. We also observed qualitatively similar types of local information transfer profiles in the PS phase, since the spatio-structure is characterized by the fixed structure of the clusters after being selected.
In BD, we can confirm that the Brownian motion of defect is clearly filtered out in SLA and SLC  (Fig. \ref{local_phase1} (c), (d); right row).  In this phase, since the background cells, which are not in defect, form a cluster of size one and each cluster reveals oscillatory behavior with almost the same frequency and with the opposite periodic phase, almost no local information transfer can be found in the background. (Note that, as explained in section \ref{sec1}, if the dynamics of the sender and the receiver are synchronized, then the information transfer reveals 0. This is because the degree of predictability of the receiver dynamics will not improve by knowing the sender dynamics.) However, the results suggest that the Brownian motion of defect receives information from the neighboring cells, meaning that the degree of predictability of the receiver dynamics (defect cells) improves by knowing the neighboring sender dynamics. From this local information transfer profile, we can infer the existence of an information transfer substrate that travels around the space.

Figure \ref{local_phase2} shows the results for SLA and SLC in DT and FDC.
In DT, similarly to the BD case, we can confirm the spatial traveling motion of defect turbulence is clearly filtered in SLA and SLC  (Fig. \ref{local_phase2} (c), (d); left row). This is caused by the same reason as the BD case. From this local information transfer profile, we can infer that the information transfer substrate travels around the space in a more complicated manner than in the BD case. In FDC, from the spatiotemporal profile of SLA and SLC, we can see that the high and low local information transfer are randomly distributed in the space, suggesting that there is no coherent structure in the dynamics (Fig. \ref{local_phase2} (c), (d); right row).

\begin{figure}[htbp]
	\centerline{\includegraphics[width=5.0in, bb=19 273 575 747]{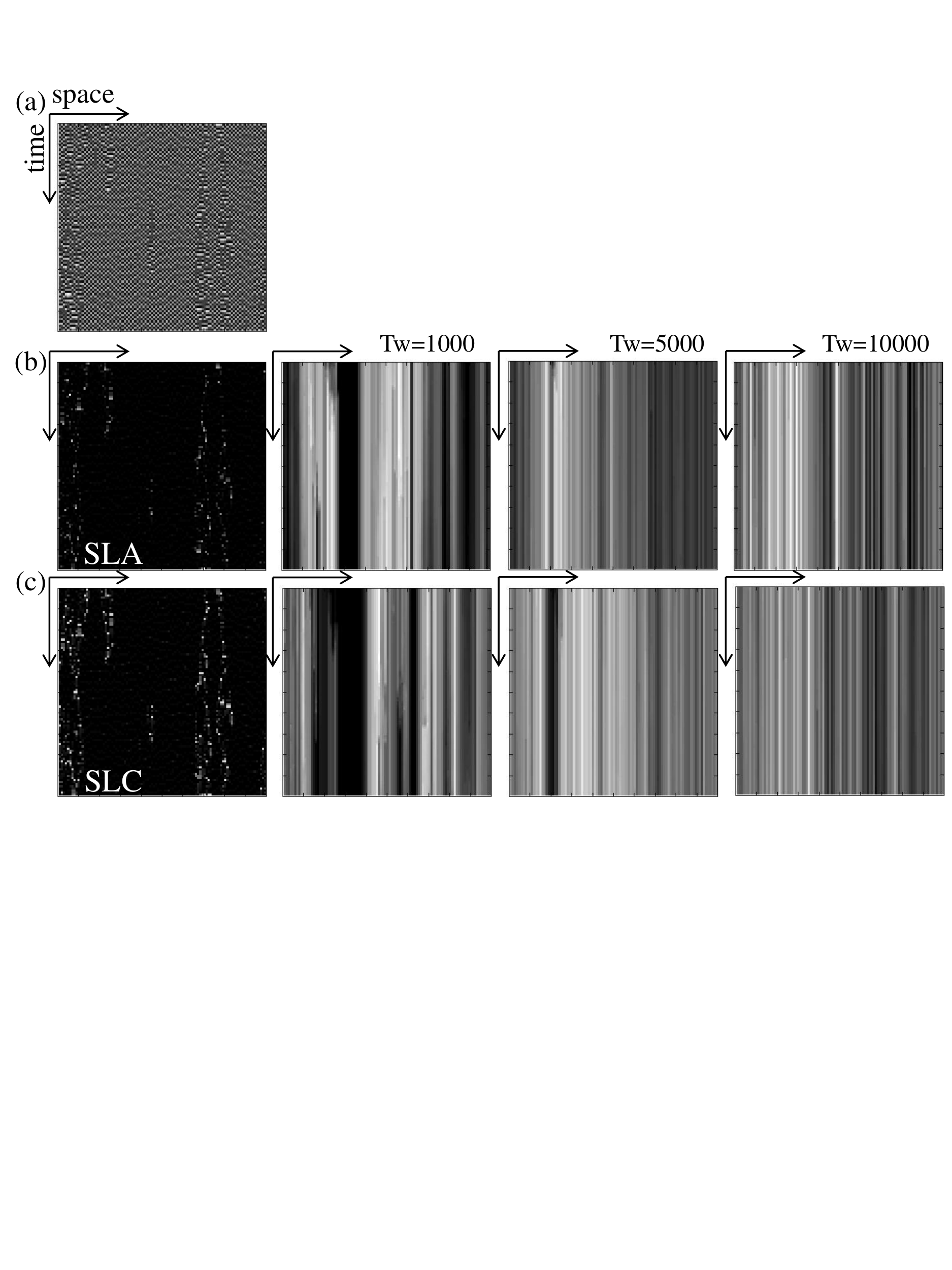}}
	\caption{Comparisons of SLA and SLC with the value of $TERV^{1}_{sa, i}$ and $TERV^{1}_{sc, i}$ calculated for each cell using the sliding window method. (a) The reference spatiotemporal dynamics with $a=1.89$ and $c=0.1$. The system was run with the same setting in Fig. \ref{local_phase2}. (b) The results of the spatiotemporal profile revealed by SLA (left figure) and the $TERV^{1}_{sa, i}$ for each cell using the sliding window method. The length of the time window is varied as $Tw = 1000, 5000,$ and $10000$. (c) The results of the spatiotemporal profile revealed by SLC (left figure) and the $TERV^{1}_{sc, i}$ for each cell using the sliding window method. The length of the time window is varied as $Tw = 1000, 5000,$ and $10000$.}
	\label{timewindow}
\end{figure}

From these results, we can suggest that the Brownian motion of defect and defect turbulence observed in the CML are the information transfer substrate of the system in Lizier's sense. Interestingly, this type of spatiotemporal profile cannot be revealed with the conventional sliding window method. Figure \ref{timewindow} shows the example of the spatiotemporal profile revealed by the sliding window method by varying the length of the time window ($Tw = 1000, 5000,$ and $10000$) in the DT phase compared with the local information transfer profile. 
For the comparison, we here simply defined measures which are $TERV$ versions of SLA and SLC as follows:
\begin{align}
TERV^{1}_{a, (i-j) \rightarrow i} &= \sum_{ \hat{x}_{i, t+1}, \hat{x}_{i, t}^{(K)}, \hat{x}_{i-j, t}^{(L)} } p(\hat{x}_{i, t+1}, \hat{x}_{i, t}^{(K)}, \hat{x}_{i-j, t}^{(L)} ) \log \frac{p(\hat{x}_{i, t+1} | \hat{x}_{i, t}^{(K)}, \hat{x}_{i-j, t}^{(L)} )}{p(\hat{x}_{i, t+1} | \hat{x}_{i, t}^{(K)} )}, \\
TERV^{1}_{c, (i-j) \rightarrow i} &= \sum_{ \hat{x}_{i, t+1}, \hat{x}_{i, t}^{(K)}, \hat{x}_{i-j, t}^{(L)},  \hat{x}_{i+j, t}^{(L)}} p(\hat{x}_{i, t+1}, \hat{x}_{i, t}^{(K)}, \hat{x}_{i-j, t}^{(L)}, \hat{x}_{i+j, t}^{(L)} ) \log \frac{p(\hat{x}_{i, t+1} | \hat{x}_{i, t}^{(K)}, \hat{x}_{i-j, t}^{(L)}, \hat{x}_{i+j, t}^{(L)} )}{p(\hat{x}_{i, t+1} | \hat{x}_{i, t}^{(K)} , \hat{x}_{i+j, t}^{(L)})},\\
TERV^{1}_{sa, i} &= TERV_{a, (i-j) \rightarrow i}^{1} + TERV_{a, (i+j) \rightarrow i}^{1}, \\
TERV^{1}_{sc, i} &= TERV_{c, (i-j) \rightarrow i}^{1} + TERV_{c, (i+j) \rightarrow i}^{1}.
\end{align}
In the plots for the sliding window method, we calculated $TERV^{1}_{sa, i}$ and $TERV^{1}_{sc, i}$ (with $j=1$) of each cell $i$ by using the given $Tw$. We can clearly see that the sliding window method shows a qualitatively different spatiotemporal profile, which does not give any information about the defect turbulence (Fig. \ref{timewindow}). This would be caused by the difference in the method for generating the probability distribution. In the conventional sliding window method, the analysis focuses on only a single cell (and its neighboring cells), so it cannot form consistent information among the overall spatiotemporal cells.

\begin{figure}[htbp]
	\centerline{\includegraphics[width=5.0in, bb=38 241 571 764]{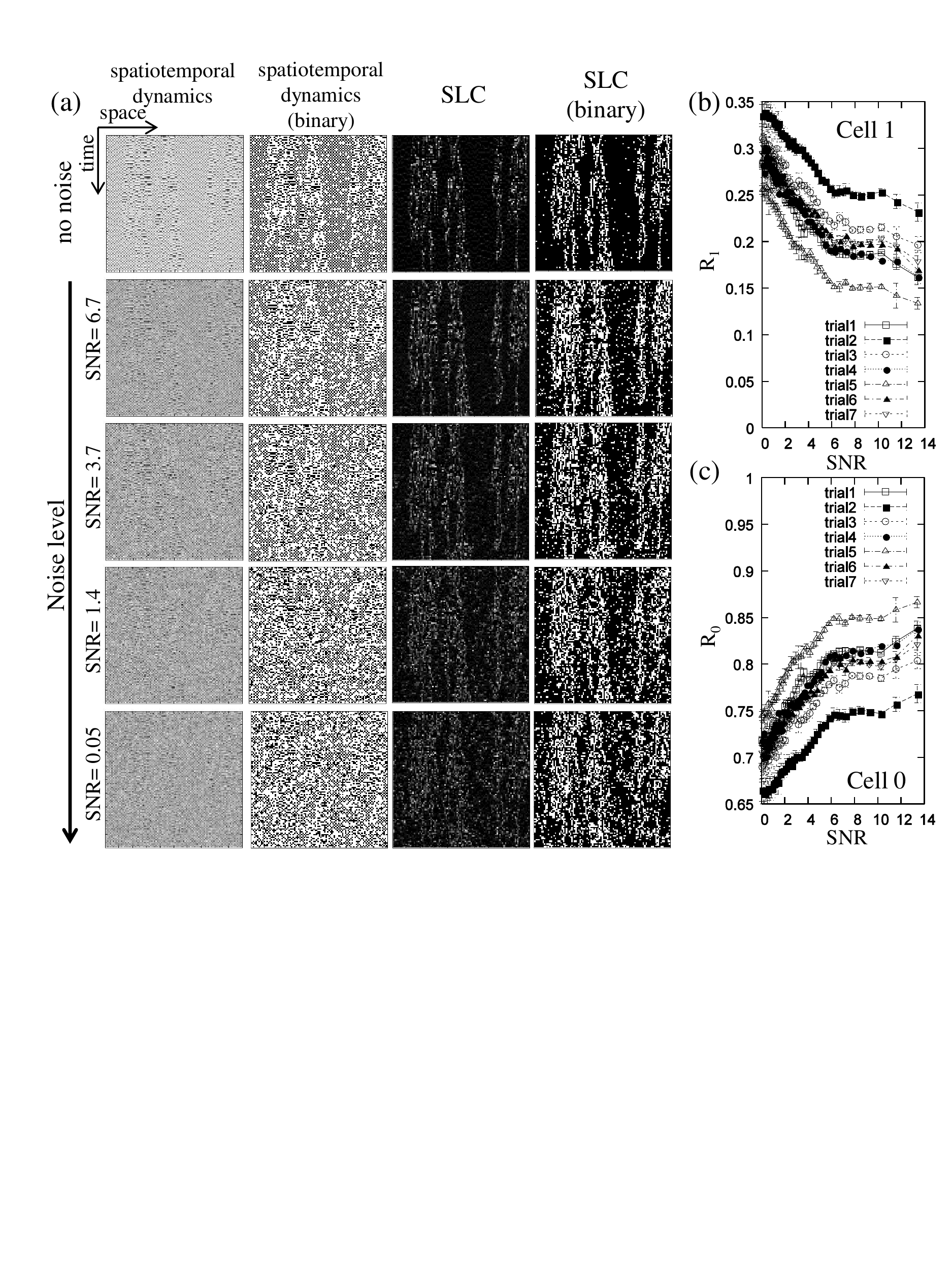}}
	\caption{(a) Typical example of spatiotemporal profiles with Gaussian noise. The figures on the left, middle left, middle right, and right show the profiles of the reference spatiotemporal dynamics ($(a, c) = (1.89, 0.1)$), the expression in binary states, the corresponding SLC profiles, and the expression in binary states, respectively. The first upper line shows the profiles without the noise, and as the line goes lower, the strength of the noise increases (expressed in SNR). The SLC is calculated with the same setting as in Fig. \ref{local_phase2}. For the profiles of the reference spatiotemporal dynamics and the corresponding SLC profiles, the darker the color, the lower the value. For the binary expression, the white and black cells are 1 and 0, respectively. See the text for details. (b) The plots show the averaged $R_{1}$ over 10 trials of noise additions regarding each noise level (SNR). Seven reference spatiotemporal dynamics are used for the analysis. In the default setting without noise, $R_{1}$ for trials 1-7 showed 0.12, 0.19, 0.15, 0.12, 0.10, 0.13, and 0.11, respectively. (c) The plots show the averaged $R_{0}$ calculated in the same experiment in (b). In the default setting without noise, $R_{0}$ for trials 1-7 showed 0.88, 0.81, 0.85, 0.88, 0.90, 0.87, and 0.89, respectively.}
	\label{noise_example2}
\end{figure}

The proposed measures can be also applied as a visualization tool or filter to stress the coherent structure existing in spatiotemporal dynamics.
To achieve this, by taking the SLC profiles, we adopted Otsu's thresholding method \cite{otsu} and expressed the profile in binary states. This method assumes that the image to be thresholded contains two classes of states (e.g., foreground and background) and then calculates the optimum threshold separating those two classes so that the intra-class variance is minimal \cite{otsu}. By taking the minimum and maximum values of the data obtained (in our case, 100 $\times$ 100 samples), we discretized the state into 20 bins and used Otsu's method to determine the threshold. Then, the value exceeding the threshold is expressed as 1 and 0 otherwise. Figure \ref{noise_example2} (a) (the upper line) shows the typical example of the results of this filtering. For comparison, we adopted the same thresholding procedure for filtering the raw spatiotemporal dynamics. As can be seen in the figure, the filtered image of the SLC profiles clearly stresses the complex traveling wave of the defect turbulences, while the filtered image of the reference spatiotemporal dynamics does not stress them as much.

To investigate the robustness of this method against the external noise, we applied various levels of Gaussian noise to the reference spatiotemporal dynamics and applied this procedure for each noise level. (Note that, as explained in Fig. \ref{local_phase2}, we run the CML (N=100) for 50000 timesteps and by using 49000 timesteps discard the first 1000 timesteps as washout to prepare the probability distributions and the final 100 timesteps are used for the SLC analysis. The Gaussian noise is added to the raw spatiotemporal dynamics for the overall 50000 timesteps for each noise level.) 
The noise level is expressed in terms of the signal-to-noise ratio (SNR [db]) as: $SNR = 10 \log_{10} (Q_{dynamics}/Q_{noise})$, where $Q_{dynamics}$ and $Q_{noise}$ are the root mean square amplitudes of the reference spatiotemporal dynamics and the Gaussian noise, respectively. Figure \ref{noise_example2} (a) (the lower lines) shows the typical filtered image with Gaussian noises. We can clearly see that the reference spatiotemporal dynamics and its corresponding filtered image starts to be buried in the noise according to the increased noise level, while the SLC profiles and their filtered images moderately sustain the foreground-background separations even if the reference spatiotemporal dynamics is almost completely buried in the noise (SNR=0.05). We also picked up 7 different reference spatiotemporal dynamics, and analyzed the transitions of the ratio of state 1 ($R_{1}$) and 0 ($R_{0}$) in the filtered SLC image over all the spatiotemporal cells. (Note that sometimes a few defect turbulences appear in the last 100 $\times$ 100 spatiotemporal cells, so we selected the spatiotemporal dynamics, which seemed to contain relatively many defect turbulences for the clarity of the analyses.) Figure \ref{noise_example2} (b) and (c) show the results. We can see that according to the increase in the noise level, $R_{1}$ ($R_{0}$) starts to increase (decrease). This was mainly driven by the increase in the state flips from 0 to 1 in the background cells. However, we can infer from the plots that, in each case, even if the reference spatiotemporal dynamics is completely buried in the noise (around SNR=0), it moderately keeps the foreground-background separations, since if the separation vanished, $R_{1}$ ($R_{0}$) should show 0.5. 

As a summary, we have demonstrated that the symbolic local transfer entropies can characterize the specific spatiotemporal profile of the CML in each phase, and especially found that the Brownian motion of defect and defect turbulence can be clearly stressed out from the background. Furthermore, we have shown that the spatiotemporal profiles revealed in these measures are qualitatively different from the profiles revealed by the conventional sliding window method, and demonstrated that these measures are robust against external noise.    

\subsection{Symbolic local information transfer in the BS-model}
\label{bsmodel}

In this section, we investigate how symbolic local information transfer is related to 
the dynamical features of the elements involved in a spatiotemporal dynamics. 
Here, we focus on the BS-model of biological evolution \cite{Bak1993,Flyvbjerg1993} 
for its simplicity. 

The original BS-model consists of $N$ species arranged on 
a one-dimensional lattice with the periodic boundary condition \cite{Bak1993}. 
Each species $i$ has its own barrier $b_i$ representing its stability or fitness. 
$b_i$ is initially given as a random number uniformly distributed in the unit interval $[0,1]$. 
At each timestep, the species $i$ that has the smallest $b_i$ is selected and its barrier 
is mutated by a new random number in $[0,1]$. This can be regarded as either a mutation of the 
species or the substitution by a new species. At the same time, the barriers of the 
nearest neighboring species are also changed to new random numbers in $[0,1]$. This latter step 
represents interaction between the species. It is well known that the BS-model exhibits a self-organized 
critical state in which all mutations occur below a critical value of $b_c \approx 0.67$ in the limit $N \to \infty$. 
The distribution of the avalanche size, which is defined as the number of consecutive mutations below 
the given threshold $b<b_c$, in the critical state shows a power law. 

\begin{figure}[htbp]
	\centerline{\includegraphics[width=5.0in, bb=49 254 535 741]{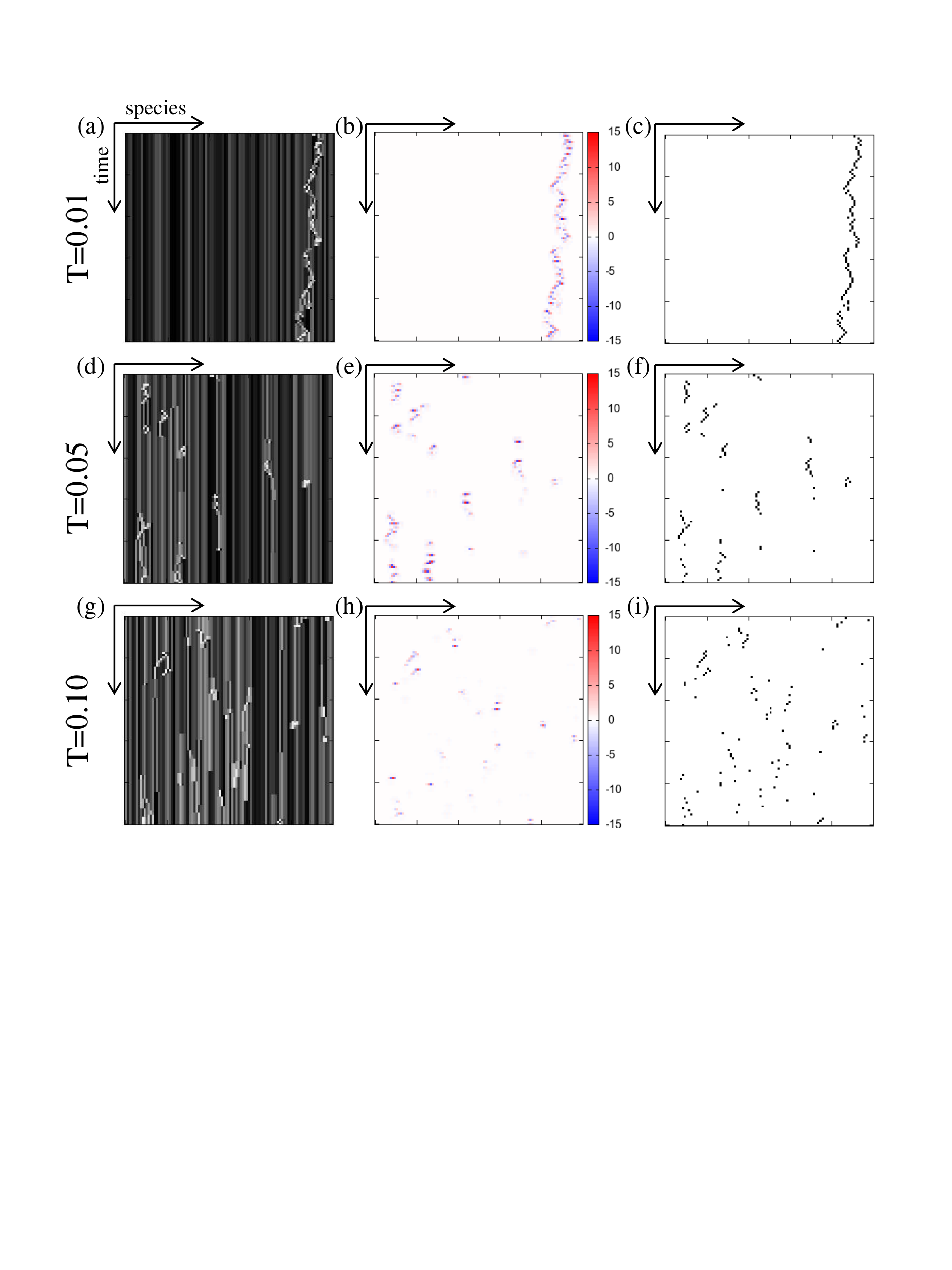}}
	\caption{For each effective temperature $T=0.01,0.05,0.10$, the spatiotemporal pattern of the barrier dynamics ((a) ,(d), (g)), 
the symbolic local information transfer difference profile ((b), (e), (h)) and the spatiotemporal positions of the 
mutations ((c), (f), (i)) are shown.}
	\label{bar-lit-act}
\end{figure}

Here, we employ a different version of the BS-model with intrinsic noise \cite{Sneppen1995} characterized by 
an effective temperature $T$. The only point that differs from the original model is that the 
selected species is not necessarily the one with the smallest barrier; however, any species with the barrier $b_i$ 
can be selected with the probability proportional to $\exp(-b_i/T)$. Thus, as $T \to 0$ we recover the 
original BS-model. In Figure \ref{bar-lit-act} (a), (d) and (g), we show the spatiotemporal behavior of 
the BS-model with $N=100$ and effective temperatures $T=0.01,0.05,0.10$, respectively. As references, 
the positions of the mutations are also shown in Figure \ref{bar-lit-act} (c), (f) and (i). 
Here, timesteps are simply measured by the number of mutations. For low $T$, mutations are consecutively clustered. 
However, the frequency of the unclustered mutations, namely, rather isolated mutations, becomes high as $T$ increases. 

In the following, we apply the symbolic local apparent transfer entropy ($ste_{a}(i, j, t+1)$, Eq. (\ref{stea})) from species $i-j$ to species $i$ at timestep $t$ by setting the embedding vectors as $(K, L) = (4, 4)$. However, we note that we obtained similar results for other values of $(K, L)$ (data not shown). 
The conditional probabilities are estimated from 10000 timesteps after the initial transient 1000 timesteps with $N=100$. 
An additional 1000 timesteps are used to study the relationship between the symbolic local information transfer 
and the stability of species below. 

We introduce the \textit{symbolic local transfer entropy difference} on species $i$ by 
\begin{equation}
\Delta te=te_{\textrm{in}} - te_{\textrm{out}}, 
\label{eq2}
\end{equation}
where 
\begin{align}
te_{\textrm{in}} &=  ste_{a}(i, 1, t+1) + ste_{a}(i, -1, t+1) \\
&= ste_{sa}(i, t+1)
\label{eq3}
\end{align}
and 
\begin{equation}
te_{\textrm{out}} =  ste_{a}(i-1, -1, t+1) + ste_{a}(i+1, 1, t+1). 
\label{eq4}
\end{equation}
Namely, $\Delta te$ on a species is the difference between the inflow and outflow of the symbolic local information transfer from/to 
the nearest neighboring species. 
In Figure \ref{bar-lit-act} (b), (e) and (h), the value of $\Delta te$ at each spatiotemporal position is shown 
for the spatiotemporal dynamics in (a), (d) and (g), respectively. From the low temperature case, we can see that 
$\Delta te$ correctly filters out the mutation sites. If we observe the high temperature case carefully, $\Delta te$ values 
at unclustered mutations are rather suppressed. Thus, it actually filters out the consecutively clustered mutations. 

\begin{figure}[htbp]
	\centerline{\includegraphics[width=5.0in, bb=38 199 557 731]{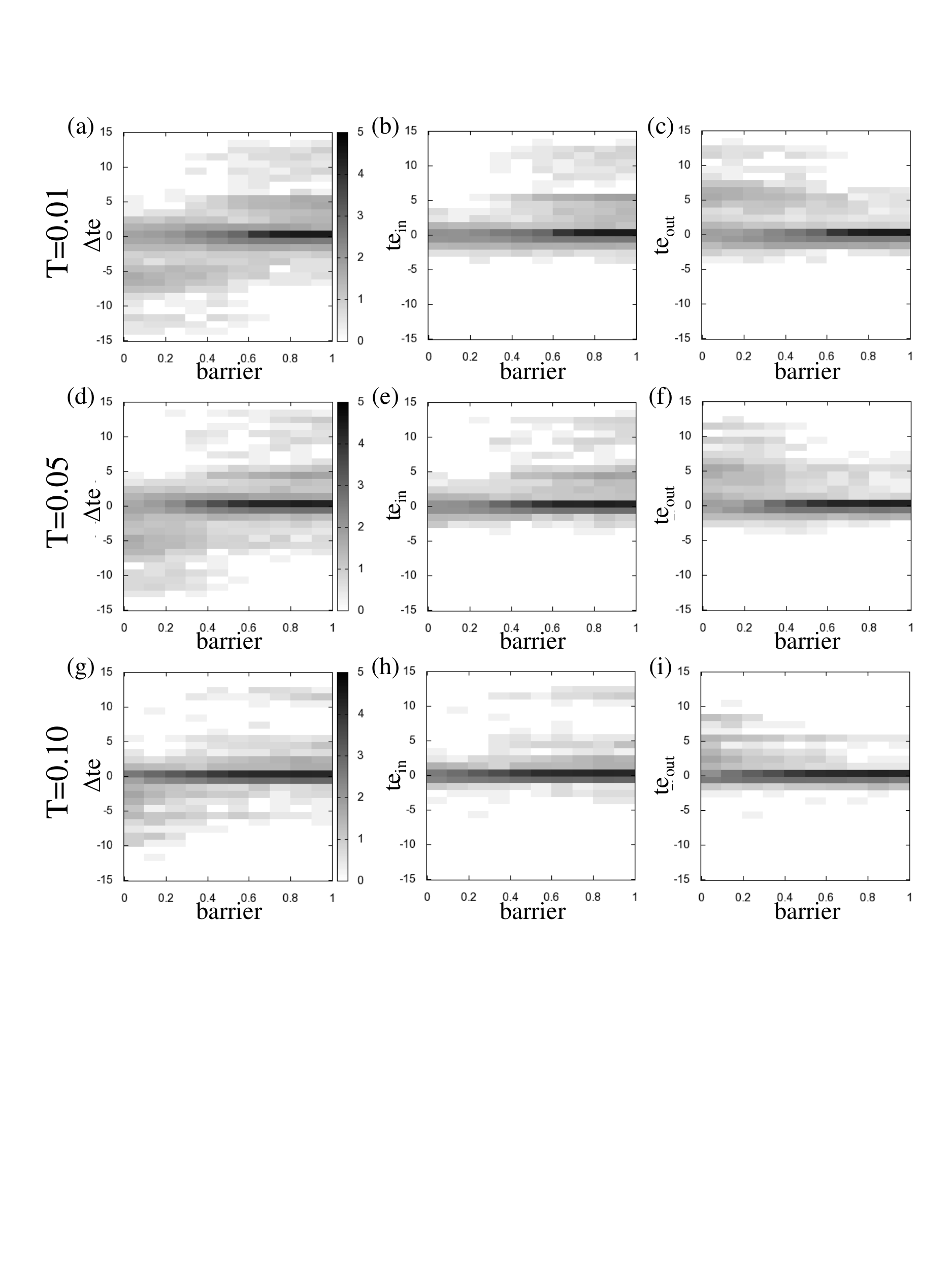}}
	\caption{The relationship between $b_i$ and $\Delta te$ ((a), (d), (g)), that between $b_i$ and $te_{\textrm{in}}$ 
((b), (e), (h)) and that between $b_i$ and $te_{\textrm{out}}$ ((c), (f), (i)) are shown for each 
effective temperature $T=0.01,0.05,0.10$. The level of grey scale 
indicates the value of $\log_{10} (1+F_{x,y})$, where $F_{x,y}$ is the frequency of the spatiotemporal positions located on 
the $b_i$-$\Delta te$ plane, such that $x \leq b_i <x+0.1$ and $y \leq \Delta te <y+1$ 
($x=0,0.1,\cdots,0.9$ and $y=-15,-14,\cdots,14$) for the left-most figures. The grey scales for the other 
figures are defined similarly.}
	\label{bar-dt}
\end{figure}

We plot the relationship between $b_i$ and $\Delta te$ in Figure \ref{bar-dt} (a), (d) and (g) for effective 
temperatures $T=0.01,0.05,0.10$, respectively. In any case, large positive $\Delta te$ (in short, $\Delta te >> 0$) 
can occur only when $b_i$ takes a high value, and a large negative $\Delta te$ (in short, $\Delta te << 0$) 
is realized only if $b_i$ is small, although these tendencies become obscure as $T$ increases. 
In other words, a large positive symbolic local information transfer difference on 
a species implies that the species is in a rather stable state. On the other hand, a large negative $\Delta te$ 
implies the species is in a rather unstable state. In the second and third 
columns of Figure \ref{bar-dt}, we also show the relationship between $b_i$ and $te_{\textrm{in}}$, and between $b_i$ and $te_{\textrm{out}}$, respectively. Although $ste_{a}(i, j, t+1)$ can take a negative value a priori, 
these figures indicate that this does not contribute significantly to large $|\Delta te|$ values. 
Namely, $\Delta te >>0$ results because of a large inflow of $te_{\textrm{in}}$ to the species, and 
$\Delta te <<0$ because of a large outflow of $te_{\textrm{out}}$ from the species. 

\begin{table}\label{tab1}
\centering
\caption{}
\begin{tabular}{lllll}
\hline\noalign{\smallskip}
$T$ & $\mu$ & $\sigma$ & $f_{\Delta te>\mu+\sigma}^{d=1}$ & $f_{\Delta te<\mu-\sigma}^{d=0}$ \\
\noalign{\smallskip}\hline\noalign{\smallskip}
0.01 & 0.04 & 9.58 & 0.99 & 1.00 \\
0.05 & 0.03 & 8.23 & 0.98 & 1.00 \\
0.10 & 0.01 & 3.87 & 0.91 & 0.67 \\
\noalign{\smallskip}\hline
\end{tabular}
\end{table}

In Table 1, we show the average and the standard deviation of $\Delta te$ over the spatiotemporal positions 
with $d=0,1,2$, where $d$ is the spatial distance on the one-dimensional array from the mutated species for $T=0.01, 0.05, 0.10$. 
$f_{\Delta te>\mu+\sigma}^{d=1}$, the fraction of the spatiotemporal positions with $d=1$ among those with $\Delta te>\mu+\sigma$, 
and $f_{\Delta te<\mu-\sigma}^{d=0}$, the fraction of the spatiotemporal positions with $d=0$ among those with $\Delta te<\mu-\sigma$, 
are also shown. If we interpret the condition $\Delta te>>0$ with $\Delta te>\mu+\sigma$ and 
$\Delta te<<0$ with $\Delta te<\mu-\sigma$, then we can say that in most cases $\Delta te>>0$ occurs only when the species is 
involved with the mutation at one of its two neighboring species, and in many cases $\Delta te<<0$ occurs only when 
the species itself is mutated. A large positive symbolic local information transfer indicates the change in a stable entity 
interacting with an unstable element. 

The above observations can be roughly explained as follows: if the value of the barrier is high 
for a species, then its change is an unexpected event that typically occurs due to interaction between the species. 
If the barrier value of the species decreases, then the permutation $\hat{b}_{i,n+1}^{(L+1)}$ 
involving the information on both the past and future states of the species changes from rare to typical. 
Hence, with possibly some subtle extra conditions, 
the information on the states of neighboring species is expected to contribute to improving the prediction of 
$\hat{b}_{i,n+1}^{(L+1)}$, namely, $te_{\textrm{in}}$ will take a large value. From the opposite standpoint, 
a species with a small barrier tends to mutate easily, and if the interaction leads to the change in the 
permutation type of the barrier sequence in one of the two neighboring species from typical to rare, 
then the information on the species will contribute to the prediction on the neighboring species, which leads 
to a large $te_{\textrm{out}}$ value.

\section{Conclusion and discussion}
\label{sec6}
In this paper, we proposed measures, symbolic local transfer entropies, to reveal the local profile existing in spatiotemporal dynamics. We then applied these measures to two test models, the CML system and BS-model, to show their relevance to spatiotemporal systems that have continuous states.
As a result, in the CML, we demonstrated that these measures can be successfully used as a spatiotemporal filter to stress a coherent structure buried in the system. In particular, we showed that the approach can clearly stress out defect turbulences or Brownian motion of defects from the background, which provides quantitative evidence that these moving patterns are the information transfer substrate in the spatiotemporal system. We then showed that these measures reveal qualitatively different properties than the conventional approach using the sliding window method, and are also robust against external noise. In the BS-model, we demonstrated that these measures can provide novel insight to the model, explaining how symbolic local information transfer is related to the dynamical properties of the elements involved in the spatiotemporal dynamics.

The methods we proposed in this paper can be extended in several ways. For example, as explained in sections \ref{sec1} and \ref{sec3}, although we focused on the special case of TERV that has $M=1$ for simplicity, it would be worth investigating how the performance of the measures differs according to the selection of $M$. In addition, several measures have recently been proposed that can capture directional and dynamic information transfer in the realm of permutation partitioning, such as {\it momentary sorting information} \cite{momentary} or {\it symbolic directed information} \cite{Kohei3}. Because of the specificity of each measure, we can expect variations in the performance when extended for the local scale. Furthermore, as explained in section \ref{intro}, local information transfer is one of the factors consisting the concept of information dynamics. The permutation-information theoretic approach to local information storage and local information modification can be explored in future work.

\bibliographystyle{epj}
\bibliography{Kohei_epj}
\clearpage
\clearpage
\clearpage
\clearpage
\clearpage
\clearpage
\clearpage

\end{document}